\documentclass[pra,preprint,aps,amsfonts,eqsecnum,nofootinbib,showpacs]{revtex4}
\usepackage[T1]{fontenc}
\usepackage[american]{babel}
\usepackage[tbtags]{amsmath}
\usepackage{amsopn}
\usepackage{indentfirst}
\usepackage{amsfonts}
\usepackage{graphicx}
\begin{document}
\title{Quantum mechanics of a free particle on a pointed plane revisited}
\author{K.\ Kowalski, K.\ Podlaski and J.\ Rembieli\'nski}
\affiliation{Department of Theoretical Physics, University
of \L\'od\'z, ul.\ Pomorska 149/153,\\ 90-236 \L\'od\'z,
Poland}
\begin{abstract}
The detailed study of a quantum free particle on a pointed plane is
performed.  It is shown that there is no problem with a mysterious ``quantum anticentrifugal
force" acting on a free particle on a plane discussed in a very recent paper: 
M.\ A.\ Cirone {\em et al}, Phys.\ Rev.\ A {\bf 65}, 022101 (2002), but we
deal with a purely topological efect related to distinguishing a point on
a plane.  The new results are introduced concerning self-adjoint extensions
of operators describing the free particle on a pointed plane as well as
the role played by discrete symmetries in the analysis of such extensions. 
\end{abstract}
\maketitle

\section{Introduction}
Although a free particle on a half-line is one of the first problems commonly used in the standard courses of quantum mechanics, nevertheless the fact that the properties of the barrier preventing a particle from a motion on the whole line, and existing bound states are related to self-adjoint extensions of the energy operator \cite{Bonneau,Garbaczewski} is frequently unfamiliar even to working in the field of quantum theory.  Indeed, the analysis of self-adjointness of operators is usually treated as a boring mathematical task without any reference to concrete properties of the physical system under consideration.

The motivation for this work was the clarification of details of the analysis of a quantum free particle on a pointed plane.  Indeed, the theory of such quantized system seems to be far from complete.  This observation is supported by the very recent paper \cite{Cirone} discussing a quantum free particle in two dimensions.  First of all, the authors of \cite{Cirone} seem to be unaware of the fact that they do not deal with the plane ${\bf R}^2$ but the pointed plane i.e.\ ${\bf R}^2\setminus\{0,0\}$.  Furthermore,  in spite
of the fact that one can find in \cite{Cirone} the reference to the celebrated monograph  \cite{Albeverio} on the self-adjoint extensions of symmetric operators in Hilbert space, the study of such extensions which is crucial for the identification of bound states of a quantum particle on a pointed plane is ignored in \cite{Cirone}.  

In this work the detailed analysis of a quantum free particle on a
pointed plane is performed.  As a matter of fact some aspects
of a quantum particle moving on a pointed plane in a Coulomb potential
have been already investigated by Schulte \cite{Schulte}. The properties of the resolvent of the extensions of the Hamiltonian were also discussed from the mathematical point of view \cite{Adami,Dabrowski} ({\em nota bene\/} these
works were not cited in \cite{Cirone}).  Nevertheless, the analysis of the problem performed in this paper is much more complete and physically oriented and we provide the new observations concerning the self-adjoint extensions of the
Hamiltonian for the quantum free particle on a pointed plane
especially in the context of the properties of the angular momentum
as well as the important role played by discrete symmetries such as the
time-reversal one. In particular, we have determined the energy spectrum of the bound states and their explicit connection with the extension parameters by means of much more simple methods comparing with the advanced mathematical approaches applied in \cite{Schulte,Adami,Dabrowski}.

\section{Preliminaries \label{II} }
We begin with a brief discussion of the Hamiltonian of a free particle on the plane ${\bf R}^2$. In the coordinate representation, the Hamiltonian is given by 
\[\hat{H}=-\frac{\hbar^2}{2M}\left(\partial_x^2+\partial_y^2\right)\equiv-\frac{\hbar^2}{2M}\Delta_2\]
and the corresponding Hilbert space is the space of square integrable functions $ L^2\left({\bf R}^2,d^2x\right)$. It is easy to see that the von Neumann deficiency index \cite{Albeverio,Reed,Akhiezer} of $\Delta_2$ is (0,0)  so the Hamiltonian is in this case essentially self-adjoint and positive definite as a sum of squares of the self-adjoint momentum operators.
Therefore it is translationaly invariant, its spectrum is ${\bf R}_+$ (the set of non negative real numbers)  and there are no bound states. This is a simple consequence of the topology of the plane: it is homogeneous and simply connected. A completely different situation arises if we extract from the plane ${\bf R}^2$ a single point. In this case the  translational invariance is lost and the pointed plane $\Dot{{\bf R}}^2={\bf R}^2\setminus\{0,0\}$ is infinitely connected i.e. its fundamental group is infinite cyclic group. As shown in this work, this fact has dramatic consequences for quantum mechanics on  $\Dot{{\bf R}}^2$.
Now the most natural coordinates for the study of the case with the pointed plane $\Dot{{\bf R}}^2$ are the polar coordinates. The origin \{0,0\} is identified with the hole in $\Dot{{\bf R}}^2$. We stress that in the polar coordinates the origin is a singular point (the Jacobian is going to 0 when $r\rightarrow0$) , of course this is no problem for the pointed plane $\Dot{{\bf R}}^2$ where this point is extracted. However, this can be a source of misinterpretations when polar coordinates are applied to the usual plane ${\bf R}^2$, because in such a  case we have a ``hidden'' extraction of the origin.

 The Hamiltonian for a free particle on the pointed plane $\Dot{{\bf R}}^2$  written in the polar coordinates takes the form
\begin{equation}\label{a1}
\hat{H}=-\frac{\hbar^2}{2M} \left(\frac{\partial^2}{\partial r^2}+\frac{1}{r}\frac{\partial}{\partial r}+\frac{1}{r^2}\frac{\partial^2}{\partial\varphi^2}\right).
\end{equation}
We designate the corresponding carrier Hilbert space of the square integrable functions on $\Dot{{\bf R}}^2$ by $L^2\left(\Dot{{\bf R}}^2,rdrd\varphi\right)$. This space can be represented as the  natural tensor product of the form $L^2\left({\bf R}_+,rdr\right)\otimes L^2\left(S^1,d\varphi\right)$
where the former is the Hilbert space of square integrable functions (with respect to the measure $rdr$)  on the half-axis ${\bf R}_+$ whereas $L^2\left(S^1,d\varphi\right)$ denote a Hilbert space of the square integrable functions on a circle $S^1$. We  discuss the structure of $L^2\left(S^1,d\varphi\right)$ in the next section. In view of the tensor product structure of the Hilbert space the Hamiltonian can be written as 
\begin{equation}\label{a2}
\hat{H}=-\frac{\hbar^2}{2M}\left[\left(\frac{\partial^2}{\partial r^2}+\frac{1}{r}\frac{\partial}{\partial r}\right)\otimes I+\frac{1}{r^2}\otimes\frac{\partial^2}{\partial\varphi^2}\right].
\end{equation}
Therefore  to analyze the problem of self-adjointness of $\hat{H}$, we should proceed in two steps:
\begin{description}
\item[(A) ] Consider the realizations of the  plane rotation group $SO\left(2\right)$ and discuss the self-adjointness of $\left(-\frac{\partial^2}{\partial\varphi^2}\right)$ in the corresponding space $L^2\left(S^1,d\varphi\right)$.
\item[(B) ] Find the self-adjoint extensions of $\hat{H}$ using the tensor product decomposition of $\hat{H}$ \eqref{a2}.
\end{description}

Let us start with discussion of the Hilbert space representation of the rotations on $S^1$.

\section{Rotational symmetry \label{III} }

Bearing in mind an  important role of the rotational invariance in the analysis of the problem, we first discuss the  realizations of  $SO\left(2\right)\approx U\left(1\right)$ group in a Hilbert space $\mathcal{H}$ of functions on $S^1$. Our basic requirement is that the  physical states as well as local observables (like current densities)  are periodic in the angle variable $\varphi$ with the period $2\pi$. 
Now, the most general action of $ SO\left(2\right)$ in $\mathcal{H}$ is of the following form:
\begin{equation}\label{a3}
\hat{U}_\lambda\left(\alpha\right) f\left(\varphi\right)=e^{i\lambda\alpha}f\left(\varphi+\alpha\right),
\end{equation} 
where $\lambda\in{\bf R}$. The factor $e^{i\lambda\alpha}$ appears because the covering group of $ U\left(1\right)$ is the additive group of real numbers. Taking into account \eqref{a3} we see that the above mentioned projective structure of $\mathcal{H}$ is preserved  with the period $2\pi$ if for each $f,g\in\mathcal{H}$
\begin{equation}\label{a4}
g^*\left(\varphi+2\pi\right)f\left(\varphi+\alpha+2\pi\right)=g^*\left(\varphi\right)f\left(\varphi+\alpha\right)
\end{equation}
for all $\varphi$ and $\alpha$. To fulfill the condition \eqref{a4} the vectors from $\mathcal{H}$ must be quasi periodic i.e. for each $f\in\mathcal{H}$ 
\begin{equation}\label{a5}
f\left(\varphi+2\pi\right)=e^{i\theta 2\pi} f\left(\varphi\right).
\end{equation}
for a fixed $\theta\in[0,1)$. If $\theta$ is rational, then  $f$ is periodic. Because the $ SO\left(2\right)$ invariant measure on $\mathcal{H}$ is (up to a real multiplier $\chi\left(e^{i\varphi}\right)$)  the usual Lebesgue measure $d\varphi$, therefore the scalar product can be written as\footnote{In the Hilbert space of  quasi periodic functions the scalar product can be defined as $\left(f,g\right)=\lim_{T\rightarrow\infty}\frac{1}{T}\int_0^Td\varphi f^*\left(\varphi\right)g\left(\varphi\right)$. It is easy to see that this product reduces to \eqref{a6} for a subspace determined by the condition \eqref{a5} with a fixed $\theta$.}
\begin{equation}\label{a6}
\left(f,g\right)=\frac{1}{2\pi}\int_0^{2\pi}d\varphi f^*\left(\varphi\right)g\left(\varphi\right).
\end{equation}

Notice that the transformation \eqref{a3} preserves the quasi periodicity condition \eqref{a5}. 
Hereafter we will denote the Hilbert space of the square integrable quasi periodic functions satisfying the quasi periodicity condition \eqref{a5} and with the scalar product \eqref{a6} by $L^2\left(S^1,d\varphi\right)_\theta$.

We now return to the realization \eqref{a3} of $SO\left(2\right)$. From the Stone theorem we know that $\hat{U}_\lambda\left(\alpha\right)$ is generated by a self-adjoint operator $\hat{J}_\lambda$ via the exponential map
\begin{equation}\label{a7}
\hat{U}_\lambda\left(\alpha\right)=e^{i\alpha \hat{J}_\lambda}.
\end{equation}
Taking into account of \eqref{a3} and \eqref{a7} we find that 
\begin{equation}\label{a8}
\hat{J}_\lambda=-i\frac{\partial}{\partial\varphi}+\lambda.
\end{equation}
Eqs. \eqref{a5},\eqref{a6} and \eqref{a8} taken together yield  $\left(f,\hat{J}_\lambda g\right)=\left(\hat{J}_\lambda f,g\right)$. Thus $\hat{J}_\lambda=\hat{J}_\lambda^\dagger$ and  consequently $\hat{J}_\lambda$ is essentially self-adjoint. The domain of $\hat{J}_\lambda$ contains absolutely continuous functions $f$ from $L^2\left(S^1,d\varphi\right)_\theta$ such that $\frac{\partial}{\partial\varphi}f\in L^2\left(S^1,d\varphi\right)_\theta$ too. This can be verified also by means of the standard theory of self-adjoint extensions of symmetric operators of von Neumann and Krein \cite{Albeverio,Reed,Akhiezer}. Indeed, the solutions of the equations
\begin{equation*}
\hat{J}_\lambda^\dagger f_\pm=\pm i f_\pm
\end{equation*}
i.e.
\begin{equation*}
\frac{\partial}{\partial\varphi}f_\pm=\left(-i\lambda\mp 1\right) f_\pm
\end{equation*}
are (up to normalization)
\begin{equation*}
f_\pm\left(\varphi\right)=e^{\left(-i\lambda\mp 1\right)\varphi}.
\end{equation*}
Of course $f_\pm$ do not belong to $L^2\left(S^1,d\varphi\right)_\theta$. Consequently, the deficiency index of $\hat{J}_\lambda$ is (0,0)  i.e. $\hat{J}_\lambda$ is essentially self-adjoint.

The next important question is related to the spectrum of $\hat{J}_\lambda$. The eigenvalue equation
\begin{equation}\label{a9}
\hat{J}_\lambda f_\mu=\mu f_\mu
\end{equation}
has the solutions of the form
\begin{equation}\label{a10}
f_\mu\left(\varphi\right)=e^{i\left(\mu-\lambda\right)\varphi}.
\end{equation}
Because $f_\mu\in L^2\left(S^1,d\varphi\right)_\theta$ they satisfy the quasi periodicity condition \eqref{a5} 
and consequently
\begin{equation}\label{a11}
\mu=\lambda+\theta-[\lambda+\theta]+l\equiv \epsilon+l,
\end{equation}
where $l$ is integer  and $\epsilon\in[0,1)$. The symbol $[x]$ designates the biggest integer in $x$. Notice that for $\lambda=0$, we have $\epsilon=\theta$.
Finally, we remark that the transformation
\begin{equation}
\label{a12}
V_\tau f\left(\varphi\right):=e^{i\tau\varphi}f\left(\varphi\right)\equiv\tilde{f}\left(\varphi\right),\quad \tau\in\left(0,1\right)
\end{equation}
maps unitarily $L^2\left(S^1,d\varphi\right)_\theta$ into $L^2\left(S^1,d\varphi\right)_{\tilde{\theta}}$,  where $\tilde{\theta}=\theta-\tau+h\left(\tau-\theta\right)$; here $h(x)$ is the Heaviside step function. Notice that for $\theta\neq\tilde{\theta}$ these spaces are orthogonal in the sense of the scalar product $\left(f,g\right)=\lim_{T\rightarrow\infty}\frac{1}{T}\int_0^Tf^*g$. 
Finally, consider the unitary phase operator $\hat{V}$ defined by
\begin{equation}\label{x}
\hat{V}f\left(\varphi\right):=e^{i\varphi}f\left(\varphi\right).
\end{equation}
It is evident that $\hat{V}$ preserves the quasi periodicity condition \eqref{a5} and consequently acts unitarily in $L^2\left(S^1,d\varphi\right)_\theta$. We can interpret $\hat{V}$ as a ``position'' operator on the circle. Using \eqref{a8} and \eqref{x} we get  
\begin{equation}\label{xx}
\hat{V}\hat{J}_\lambda=\left(J_\lambda-1\right)\hat{V}
\end{equation}
so $\hat{V}$ and $\hat{V}^\dagger$ are the ladder operators. Furthermore, we 
can represent $\hat{V}$ as $\hat{V}=e^{i\hat{\varphi}}$, where the 
self-adjoint angle operator $\hat{\varphi}$ is defined as
\begin{equation}\label{xxx}
\hat{\varphi}f\left(\varphi\right)=\left(\varphi-2\pi\left[\frac{\varphi}{2\pi}
\right]\right)f\left(\varphi\right)
\end{equation}
to preserve the quasi periodicity condition. Because $\hat{\varphi}$ is 
bounded therefore its domain is the whole space 
$L^2\left(S^1,d\varphi\right)_\theta$.

\section{The Operator $-\partial^2/\partial\varphi^2\equiv \hat{D}^2$}
\label{IV}

As was discussed in the previous section, to preserve periodicity of the circle $S^1$ under rotations  on the level of the Hilbert space representation, it is necessary to deal with the space of quasi periodic functions. For this reason a sensible discussion of the operator $-\frac{\partial^2}{\partial\varphi^2}$ as a self-adjoint operator with $SO\left(2\right)$ invariant domain must be given in the context of the space $L^2\left(S^1,d\varphi\right)_\theta$. Notice, that the action of $-\frac{\partial^2}{\partial\varphi^2}$ does not violate the quasi periodicity condition \eqref{a5} and consequently leaves invariant  the domain  $L^2\left(S^1,d\varphi\right)_\theta$. Taking into account the form of the scalar product \eqref{a6} and quasi periodicity condition \eqref{a5} we obtain $\left(f,\hat{D}^2 g\right)=\left(\hat{D}^{2\dagger} f,g\right)$.
 Thus $\hat{D}^{2\dagger}=\hat{D}^2$ i.e. it is essentially self-adjoint in $L^2\left(S^1,d\varphi\right)_\theta$. Indeed, it is easy to verify that the deficiency spaces for $D^{2\dagger}$ in $L^2\left(S^1,d\varphi\right)_\theta$ are zero dimensional so the corresponding deficiency index is (0,0) . The domain of $\hat{D}^2$ contains functions such that $\frac{\partial^2}{\partial \varphi^2}f\in L^2\left(S^1,d\varphi\right)_\theta$. Therefore, we have 
\begin{equation}\label{a13}
\hat{D}^2=\left(\hat{J}_\lambda-\lambda\right)^2.
\end{equation}
Consequently, the spectrum of $\hat{D}^2$ in $L^2\left(S^1,d\varphi\right)_\theta$ is of the form (see eq.\eqref{a13} and \eqref{a11}) 
\begin{equation}\label{a14}
\left(\theta+m\right)^2,\quad m\in{\bf Z},
\end{equation}
where ${\bf Z}$ designates the set of integers.
The common eigenvectors of $\hat{J}_\lambda$ and $\hat{D}^2$ (see eq. \eqref{a10})  can be rewritten in the   more convenient form with respect to $\hat{D}^2$, namely
\begin{equation}\label{a15}
f_{\theta,m}\left(\varphi\right)=e^{i\left(m+\theta\right)\varphi}.
\end{equation}
It should be noted that for $\lambda=0$ we have $m=l$, however in general $m=l-[\lambda+\theta]$.

We finally recall that in the $n>2$ dimensional case the corresponding rotation group $SO\left(n\right)$ is not infinitely many connected as it takes place in the $SO\left(2\right)$ case as well as the first fundamental group for $\dot{{\bf R}}^n$, $n>2$, is trivial contrary to the $n=2$ case. This is the reason of the qualitative distinctions between free quantum mechanics on $\dot{{\bf R}}^2$ and $\dot{{\bf R}}^n$, $n>2$.

\section{Quantum Hamiltonian for a free motion on $\dot{{\bf R}}^2$ \label{V}}  

Now we are in a position to find all self-adjoint extensions of the Hamiltonian \eqref{a2} which acts in the tensor product $L^2\left({\bf R}_+,rdr\right)\otimes L^2\left(S^1,d\varphi\right)_\theta$ with a fixed $\theta\in[0,1)$. In view of our observations from  sections \ref{III} and \ref{IV}, $L^2\left(S^1,d\varphi\right)_\theta$ is the direct sum of the one dimensional eigenspaces $\mathcal{H}_{\left(\theta,m\right)}=\text{span}\{f_{\theta,m}\left(\varphi\right)\}$ of $\hat{J}_\lambda$ and $\hat{D}^2$. Thus $\hat{H}$ acts in the direct sum of the tensor products $L^2\left({\bf R}_+,rdr\right)\otimes\mathcal{H}_{\left(\theta,m\right)}$, i.e. in
\begin{equation}\label{a16}
L^2\left({\bf R}_+,rdr\right)\otimes L^2\left(S^1,d\varphi\right)_\theta=\bigoplus_{m\in{\bf Z}}L^2\left({\bf R}_+,rdr\right)\otimes\mathcal{H}_{\left(\theta,m\right)}.
\end{equation}
Using  \eqref{a2},\eqref{a13},\eqref{a14} and \eqref{a15} we find that the Hamiltonian takes the following form in a subspace 
$L^2\left({\bf R}_+,rdr\right)\otimes\mathcal{H}_{\left(\theta,m\right)}$:
\begin{equation}\label{a17}
\hat{H}_{\theta,m}=-\frac{\hbar^2}{2M} \left(\frac{\partial^2}{\partial r^2}+\frac{1}{r}\frac{\partial}{\partial r}-\frac{\left(\theta+m\right)^2}{r^2}\right)\otimes I.
\end{equation}
Our purpose now is to determine the  possible deficiency spaces of $\hat{H}_{\theta,m}^\dagger$ in the space $L^2\left({\bf R}_+,rdr\right)$    with the scalar product
\begin{equation}\label{a18}
\left(F,G\right)=\int_0^\infty rdr F^*\left(r\right)G\left(r\right),
\end{equation}
where $F,G\in L^2\left({\bf R}_+,rdr\right)$. It is easy to check that $\hat{H}_{\theta,m}$ is symmetric for $G\left(r\right)$ such that $G'$ and $G''$ (here prime designates differentiation with respect to $r$)  also belong to $L^2\left({\bf R}_+,rdr\right)$  and satisfy the boundary condition $G\left(0\right)=G'\left(0\right)=0$. The conjugate operator $\hat{H}^\dagger_{\theta,m}$ has also the form \eqref{a17} however its domain is in general larger than $\hat{H}_{\theta,m}$. In order to find the self-adjoint extensions of  $\hat{H}$ we apply the von Neumann-Krein method. We seek the  solutions of the eigenvalue equations
\begin{equation}\label{a19}
\hat{H}_{\theta,m}^\dagger F_{\pm m}\left(r\right)=\pm i\kappa F_{\pm m}\left(r\right),
\end{equation}
where $\kappa\in{\bf R}$ is introduced to keep the dimension of the right hand side of \eqref{a19}; however as shown in the next section, it has a definite physical meaning.
From  \eqref{a17} and  \eqref{a19} it follows that
\begin{widetext}
\begin{equation}\label{a20}
F_{\pm m}\left(r\right)=C_1I_{m+\theta}\left(\frac{r}{\hbar}\sqrt{\mp i 2M\kappa}\right)+C_2K_{m+\theta}\left(\frac{r}{\hbar}\sqrt{\mp i 2M\kappa}\right),
\end{equation}\end{widetext}
where $I_\mu$ and $K_\nu$ are the modified Bessel functions and MacDonald functions, respectively.
Now, from the asymptotic behavior of $I_\mu$ and  $K_\nu$ in the regions $r\rightarrow 0$ and $r\rightarrow \infty$ (see Appendix)  we deduce that \eqref{a20} are elements of  $L^2\left({\bf R}_+,rdr\right)$ only for $\left(m+\theta\right)\in\left(-1,1\right)$. Therefore $m=0$ and $\theta\in[0,1)$ or $m=-1$ and  $\theta\in\left(0,1\right)$. Thus for $m\neq0,-1$ and for $m=-1$ and $\theta=0$ the deficiency indices are (0,0)  so $\hat{H}_{\theta,m}$ is in these cases essentially self-adjoint. On the other hand,  $\hat{H}_{0,0}$ and $\hat{H}_{\theta\neq 0,-1}$ are only symmetric because \eqref{a19} has for such $H_{\theta,m}$ the solutions
\begin{equation}\label{a21}
K_\theta\left(\frac{r}{\hbar}\sqrt{\mp i2M\kappa}\right)\quad \text{and}\quad K_{1-\theta}\left(\frac{r}{\hbar}\sqrt{\mp i2M\kappa}\right),
\end{equation}
respectively from $L^2\left({\bf R}_+,rdr\right)$.

Now, because  we are looking for the self-adjoint extensions of $\hat{H}$ we must solve the equation
\begin{equation}\label{a22}
\hat{H}\Psi_\pm=\pm i\kappa\Psi_\pm.
\end{equation}

Expanding $\Psi_\pm$ with respect to $f_{\theta,m}\left(\varphi\right)$ given by \eqref{a15}, using \eqref{a2} and the above results we find that\\\null

\underline{for $\theta=0$} we have one solution for ($+$) and one for ($-$)
\begin{equation}\label{a23}
\Psi_\pm^{\left(0,0\right)}=\frac{2}{\hbar}\sqrt{\frac{2M\kappa}{\pi}}K_0\left(\frac{r}{\hbar}\sqrt{\mp i2M\kappa}\right),
\end{equation}
\\\null

\underline{for $\theta\in\left(0,1\right)$} we have two solutions for (+)  and two for ($-$)  

\begin{equation}\label{a24}
\begin{split}
\Psi_\pm^{\left(\theta,0\right)}&=\frac{2}{\hbar}\sqrt{\frac{2M\kappa\cos\left(\frac{\theta\pi}{2}\right)}{\pi}}K_\theta\left(\frac{r}{\hbar}\sqrt{\mp i2M\kappa}\right)e^{i\theta\varphi},\\
\Psi_\pm^{\left(\theta,-1\right)}&=\frac{2}{\hbar}\sqrt{\frac{2M\kappa\sin\left(\frac{\theta\pi}{2}\right)}{\pi}}K_{1-\theta}\left(\frac{r}{\hbar}\sqrt{\mp i2M\kappa}\right)e^{i\left(\theta-1\right)\varphi}.\\
\end{split}
\end{equation}
Therefore, in the case of $\theta=0$ the deficiency index of $\hat{H}$ is (1,1) and for $\theta\neq0$ it is (2,2) . Thus, applying the von Neumann-Krein theory we arrive in the former case at the one parameter family of extensions and in the latter case the four parameter family. We remark that the parameters labelling the family of self-adjoint extensions of the Hamiltonian are usually related to the properties of a barrier. An excellent example is a particle in a box \cite{Carreau,Fulop2,Luz}. The authors do not know such relationship in the discussed highly nontrivial case of the pointed plane.

\subsection{The case $\theta=0$}
We now discuss the case with $\theta=0$ (see \eqref{a23}) . According to the von Neumann-Krein theory the domain of $\hat{H}$ contains in this case the vectors of the form
\begin{widetext} 
\begin{equation}\label{a25}
\Psi_0\left(r,\varphi\right)=\chi_0\left(r,\varphi\right) + C\left(K_0\left(\frac{r}{\hbar}\sqrt{-i 2M\kappa}\right)+e^{i\eta}K_0\left(\frac{r}{\hbar}\sqrt{i2M\kappa}\right)\right),
\end{equation}\end{widetext}
where $\chi_0\left(0,\varphi\right)=\chi'_0\left(0,\varphi\right)=0$,  $C$ is an arbitrary complex number and $\eta\in[-\pi,\pi)$ fix the domain. Of course $\chi_0\left(r,\varphi\right)\in L^2\left({\bf R}_+,rdr\right)\otimes L^2\left(S^1,d\varphi\right)_0$ in this case. In particular, $\chi_0\left(r,\varphi+2\pi\right)=\chi_0\left(r,\varphi\right)$.

\subsection{The case $\theta\in\left(0,1\right)$}
The case of $\theta\in\left(0,1\right)$ (see \eqref{a24})  is more complicated than the case with $\theta=0$ discussed above. Applying the von Neumann-Krein theory we find that the domain of $\hat{H}$ contains the vectors of the form
\begin{equation}\label{a26}
\Psi_\theta\left(r,\varphi\right)=\chi_\theta\left(r,\varphi\right)
 +\left(A,B\right)\left[\left(\begin{matrix}\Psi_+^{\left(\theta,0\right)}\left(r,\varphi\right)\\\Psi_+^{\left(\theta,-1\right)}\left(r,\varphi\right)\end{matrix}\right)+U\left(\begin{matrix}\Psi_-^{\left(\theta,0\right)}\left(r,\varphi\right)\\\Psi_-^{\left(\theta,-1\right)}\left(r,\varphi\right)\end{matrix}\right)\right],
\end{equation}
where $\chi_\theta\left(0,\varphi\right)=\chi_\theta'\left(0,\varphi\right)=0$, $\chi_\theta\left(r,\varphi\right)\in L^2\left({\bf R}_+,rdr\right)\otimes L^2\left(S^1,d\varphi\right)_\theta$ i.e. $\chi_\theta\left(r,\varphi+2\pi\right)=e^{i2\pi\theta}\chi_\theta\left(r,\varphi\right)$, $\left(A,B\right)$ is arbitrary complex two dimensional row vector, $\Psi_\pm$ are given by \eqref{a24} and $U$ is a fixed unitary $2\times 2$ matrix defining this self-adjoint extension. Therefore,  demanding the rotational invariance of the domain of $\hat{H}$ i.e. preservation of  the form of the second term in the eq. \eqref{a26}, applying $\hat{U}_\lambda\left(\alpha\right)$ (see \eqref{a3} and \eqref{a7}) to both sides of the eq.\eqref{a26} and absorbing irrelevant phases in the row $\left(A,B\right)$ we find that the matrix $U$ must be diagonal. Thus it turns out that the rotational invariance reduces the family of extensions to the two parameter one. More precisely, we can write \eqref{a26} in the form 
\begin{equation}\label{a27}
\begin{split}
\Psi_\theta\left(r,\varphi\right)=\chi_\theta\left(r,\varphi\right)&+A e^{i\theta\varphi}\left(K_\theta\left(\frac{r}{\hbar}\sqrt{-i 2M\kappa}\right)+e^{i\rho}K_\theta\left(\frac{r}{\hbar}\sqrt{i 2M\kappa}\right)\right)\\
&+B e^{i\left(\theta-1\right)\varphi}\left(K_{1-\theta}\left(\frac{r}{\hbar}\sqrt{-i 2M\kappa}\right)+e^{i\eta}K_{1-\theta}\left(\frac{r}{\hbar}\sqrt{i 2M\kappa}\right)\right),
\end{split}
\end{equation}
where $\rho,\eta\in[-\pi,\pi)$ are constants parametrizing the self-adjoint extensions of $\hat{H}$.

\section{The spectrum of $\hat{H}$ \label{VI}}
As is well known and easy to show, the spectrum of $\hat{H}$ contains continuous non negative part from 0 to infinity and possible bound states corresponding to negative energies. We now concentrate on the negative energy case. Since the Hamiltonian commute with $\hat{U}_\lambda\left(\alpha\right)$ it can be diagonalized in the negative part of its spectrum together with $\hat{U}_\lambda\left(\alpha\right)$ . Consequently, the eigenstates of $\hat{H}$ have determined value of $m$. Therefore, the eigenvalue equation for the radial part of the eigenvector of $\hat{H}$ can be written as 
\begin{equation}\label{a28}
\hat{H}_{\theta,m} \Psi_E^{\left(\theta,m\right)}\left(r\right)=-|E|\Psi_E^{\left(\theta,m\right)}\left(r\right).
\end{equation}
The general solution to \eqref{a28} is expressed (up to normalization) by the MacDonald functions

\begin{equation}
\label{a29}
\Psi_E^{\left(\theta,m\right)}\left(r,\varphi\right)=\Psi_E^{\left(\theta,m\right)}\left(r\right)e^{i\left(\theta+m\right)\varphi}=K_{\theta+m}\left(\frac{r}{\hbar}\sqrt{2M|E|}\right)e^{i\left(\theta+m\right)},
\end{equation}
where $\left(\theta+m\right)\in\left(-1,1\right)$ because $K_{\theta+m}\in L^2\left({\bf R}_+,rdr\right)$ only when this condition is valid. In the following we consider  the cases $\theta=0$ and $\theta\neq 0$ separately.

\subsection{The case $\theta=0$}
We first study the case of $\theta=0$ (so $m=0$). Since the corresponding solutions $\Psi_E^{(0,0)}$ of \eqref{a28} belong to domain of $\hat{H}$ therefore it is of the form \eqref{a25}, that is 

\begin{equation}\label{a30}
\begin{split}
\Psi_E^{\left(0,0\right)}\left(r\right)&=\chi_E^{\left(0,0\right)}\left(r\right)+C_E\left(K_0\left(\frac{r}{\hbar}\sqrt{-i 2M\kappa}\right)+e^{i\eta}K_0\left(\frac{r}{\hbar}\sqrt{i 2M\kappa}\right)\right)\\
&\equiv K_0\left(\frac{r}{\hbar}\sqrt{2 M|E|}\right),\quad \eta\in[-\pi,\pi).
\end{split}
\end{equation}

\begin{figure}[h]
		\includegraphics[scale=0.7]{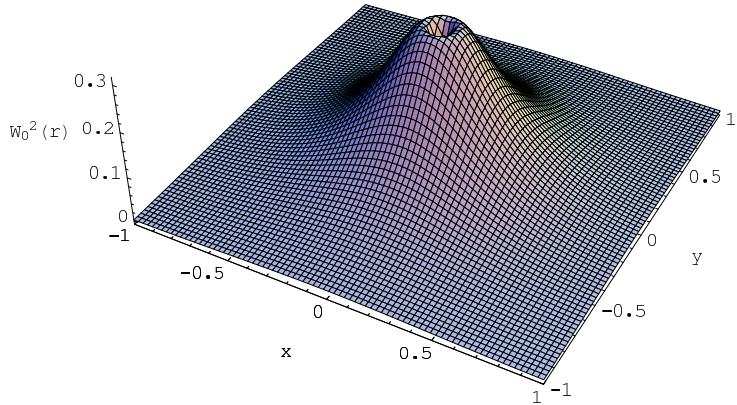}
		\caption{\label{wykK0}The probability density $W_0^2\left(r\right)=r|\Psi_E^{\left(0,0\right)}\left(r\right)|^2$ with respect to the measure $dr$ (not $rdr$), where  $\Psi_E^{\left(0,0\right)}$ is given by eq. \eqref{a29} and $r=\sqrt{x^2+y^2}$. Notice that $W_0^2\left(r\right)$ approaches zero when $r$ tends to zero.}
\end{figure}

Therefore
\begin{equation}
\chi_E^{\left(0,0\right)}\left(r\right)=K_0\left(\frac{r}{\hbar}\sqrt{2M|E|}\right)-C_E\left(K_0\left(\frac{r}{\hbar}\sqrt{-i2M\kappa}\right)+e^{i\eta}K_0\left(\frac{r}{\hbar}\sqrt{i2M\kappa}\right)\right),
\end{equation}
where  $\chi_E^{\left(0,0\right)}\left(0\right)=0$ and $\left. \frac{\partial}{\partial r}\chi_E^{\left(0,0\right)}\left(r\right)\right|_{r=0}=0$. If we apply the last condition satisfied by $\chi^{\left(0,0\right)}\left(r\right)$ by taking the derivative of both sides of \eqref{a30} with respect to $r$ and make use of some elementary properties of the MacDonald functions (see Appendix)  we get
\begin{equation}\label{a31}
E=-\kappa e^{-\frac{\pi}{2}\tan{\frac{\eta}{2}}},
\end{equation}
\begin{equation}\label{a32}
C_E=\frac{1}{1+e^{i\eta}}.
\end{equation}

\begin{figure}[h]
		\begin{center}
		\includegraphics[scale=0.5]{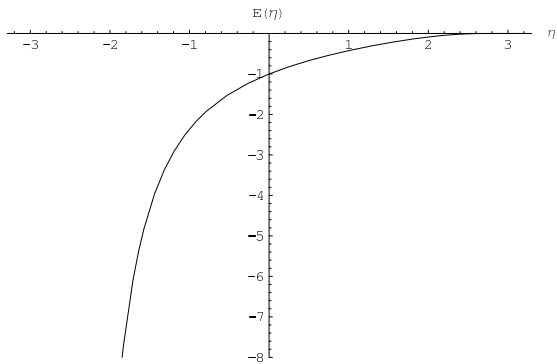}
		\end{center}
		\caption{\label{wykE0}Energy of the bound state \eqref{a31}. We point out that we have one-to-one correspondence between $\eta$ and the energy $E(\eta)$ of the bound state. Therefore $E$ can be treated as the  extension parameter instead of $\eta$.}
\end{figure}

In view of \eqref{a31} it is clear that the parameter $\kappa$ fixes the energy scale. Moreover, for a fixed self-adjoint extension given by the  concrete $\eta$ we have only one bound state. 
Notice that for $\eta=-\pi$ we have singularity in the formula \eqref{a31} and \eqref{a32} i.e. there is no finite energy bound state for $\eta=-\pi$.

\subsection{The case $\theta\in\left(0,1\right)$}
We now investigate the case with $\theta\in\left(0,1\right)$. In this case $m$ takes the values 0 or -1.
\begin{itemize}
\item\underline{For $m=0$} we have from \eqref{a27} and \eqref{a29} 

\begin{equation}\label{a33}
\begin{split}
\Psi_{E_{0}}^{\left(\theta,0\right)}\left(r,\varphi\right)&=\chi_{E_{0}}^{\left(\theta,0\right)}\left(r\right)e^{i\theta\varphi}+A_{E_{0}}e^{i\theta\varphi}\left[K_\theta\left(\frac{r}{\hbar}\sqrt{-i 2M\kappa}\right)+ e^{i\rho}K_\theta\left(\frac{r}{\hbar}\sqrt{i 2M\kappa}\right)\right]\\&\equiv K_\theta\left(\frac{r}{\hbar}\sqrt{2M|E_0|}\right)e^{i\theta\varphi}.
\end{split}
\end{equation}
By applying the same procedure as before we find that
\begin{equation}\label{a34}
E_0=-\kappa\left(\frac{\cos\left(\frac{\rho}{2}+\theta\frac{\pi}{4}\right)}{\cos\left(\frac{\rho}{2}-\theta\frac{\pi}{4}\right)}\right)^\frac{1}{\theta}
\end{equation}
and the solution \eqref{a34} to \eqref{a33} exists only for
\begin{equation}\label{a35}
\rho\in\left(-\pi+\frac{\theta\pi}{2},\pi-\frac{\theta\pi}{2}\right].
\end{equation}

The constant $A_{E_{0}}$ is given by
\begin{equation}\label{a36}
A_{E_{0}}=-\frac{e^{-i\frac{\rho}{2}}}{\sqrt{2\left(\cos\rho+\cos\left(\theta\frac{\pi}{2}\right)\right)}}.
\end{equation}

\item\underline{For $m=-1$} we obtain from \eqref{a27} and \eqref{a29} 
 
\begin{equation}\label{a37}
\begin{split}
\Psi_{E_{-1}}^{\left(\theta,-1\right)}\left(r,\varphi\right)&=\chi_{E_{-1}}^{\left(\theta,-1\right)}\left(r\right)e^{i\left(\theta-1\right)\varphi}\\
&+B_{E_{-1}}e^{i\left(\theta-1\right)\varphi}\left[K_{1-\theta}\left(\frac{r}{\hbar}\sqrt{-i 2M\kappa}\right)+ e^{i\eta}K_{1-\theta}\left(\frac{r}{\hbar}\sqrt{i 2M\kappa}\right)\right]\\&\equiv K_{1-\theta}\left(\frac{r}{\hbar}\sqrt{2M|E_{-1}|}\right)e^{i\left(\theta-1\right)\varphi}.
\end{split}
\end{equation}
The same procedure as before leads to 
\begin{equation}\label{a38}
E_{-1}=-\kappa\left(\frac{\cos\left(\frac{\eta}{2}+\left(1-\theta\right)\frac{\pi}{4}\right)}{\cos\left(\frac{\eta}{2}-\left(1-\theta\right)\frac{\pi}{4}\right)}\right)^\frac{1}{1-\theta},
\end{equation}
\begin{equation}\label{a39}
B_{E_{-1}}=\frac{e^{-i\frac{\eta}{2}}}{\sqrt{2\left(\cos\eta+\sin\left(\theta\frac{\pi}{2}\right)\right)}}.
\end{equation}
The solutions \eqref{a38} exist only for 
\begin{equation}\label{a40}
\eta\in\left(-\left(1+\theta\right)\frac{\pi}{2},\left(1+\theta\right)\frac{\pi}{2}\right].
\end{equation}
\end{itemize}

In summary, in the case of $\theta\neq0$ we have three possibilities:
\begin{itemize}
\item there are no bound states when $\rho$ and $\eta$ do not satisfy \eqref{a35} and \eqref{a40};
\item there is one bound state if only one from  the parameters $\rho,\eta$ satisfy \eqref{a35} or \eqref{a40};
\item there are two bound states if both $\rho$ and $\eta$ satisfy \eqref{a35} or \eqref{a40}.
\end{itemize}
\section{The time reversal symmetry } 
In this section we analyze the role of the time reversal symmetry. More precisely, we show that such symmetry which is most natural for the discussed case of a free dynamies considerably reduce the family of possible realizations of quantum mechanics on $\Dot{{\bf R}}^2$.

The operator of time inversion  must be antiunitary to preserve the canonical structure of quantum mechanics. In our case its action on the wave functions $\psi\left(r,\varphi\right)$ is given by the following formula \cite{Ballentine}:
\begin{equation}\label{a41}
\hat{T}\psi\left(r,\varphi\right)=\xi\psi^*\left(r,\varphi\right),
\end{equation}
where $\xi$ is a fixed phase i.e. $|\xi|^2=1$. The {\em sine qua non} condition to discuss the role of $\hat{T}$ is its existence in the Hilbert space under consideration that is  in $ L^2\left({\bf R}_+,rdr\right)\otimes L^2\left(S^1,d\varphi\right)_\theta$.
Therefore, if $\hat{T}$ symmetry is required, the domain of $\hat{H}$ should be invariant under the action of $\hat{T}$.

It is obvious that in order to define $\hat{T}$ in the above product space it is enough to define it in the spaces $ L^2\left({\bf R}_+,rdr\right)$ and $ L^2\left(S^1,d\varphi\right)_\theta$ separately. In $ L^2\left({\bf R}_+,rdr\right)$ the action \eqref{a41} is well defined because it  does not affect the asymptotic behavior of vectors in the spatial infinity ($r\rightarrow \infty$) . Nevertheless, in $ L^2\left(S^1,d\varphi\right)_\theta$ the situation is  different. Applying $\hat{T}$ to the defining relation \eqref{a5} for the quasi periodic functions we get
\begin{equation}\label{a42}
\hat{T}f\left(\varphi+2\pi\right)=e^{-i2\pi\theta}\hat{T}f\left(\varphi\right)
\end{equation}
following from  the antiunitarity of $\hat{T}$. Consequently, the conditions \eqref{a5} and \eqref{a42} are compatible only for $\theta=0$ or $\theta=\frac{1}{2}$ i.e. for periodic and antiperiodic functions with the period $2\pi$. Therefore, the time inversion operation can be defined only for these two cases i.e. in the spaces $ L^2\left({\bf R}_+,rdr\right)\otimes L^2\left(S^1,d\varphi\right)_0$ or $ L^2\left({\bf R}_+,rdr\right)\otimes L^2\left(S^1,d\varphi\right)_{\frac{1}{2}}$ 
Now, let us analyze the invariance of the domain of the Hamiltonian $\hat{H}$ in these two cases.\\\null

\underline{The case $\theta=0$}\\
By applying  $\hat{T}$ to both sides of  \eqref{a25}, using $\left(K_0\left(z\right)\right)^*=K_0\left(z^*\right)$, absorbing a phase in $C^*$ and taking into account that the complex conjugation does not change the boundary conditions for $\chi_0\left(r,\varphi\right)$, we again obtain for $\hat{T}\Psi_0$ the same relation \eqref{a25}. Thus the domain of $\hat{H}$ is in this case $\hat{T}$-invariant.
\\\null

\underline{The case $\theta=\frac{1}{2}$}\\
By applying  $\hat{T}$ to both sides of \eqref{a27} we find that  $\hat{T}\psi_{\frac{1}{2}}$ satisfy the same form of \eqref{a25} if $\rho=\eta$ i.e. we arrive at the following family of extensions:
\begin{equation}\label{a43}
\Psi_{\frac{1}{2}}\left(r,\varphi\right)=\chi_{\frac{1}{2}}\left(r,\varphi\right)+\left(Ae^{i\frac{\varphi}{2}}+Be^{-i\frac{\varphi}{2}}\right)\frac{e^{-\frac{\sqrt{M\kappa}}{\hbar}r}}{\sqrt{r}}\left(e^{i\frac{\sqrt{M\kappa}}{\hbar}r}+e^{i\left(\eta-\frac{\pi}{4}\right)}e^{-i\frac{\sqrt{M\kappa}}{\hbar}r}\right),
\end{equation}
where we have used the explicit form of the function $K_{\frac{1}{2}}\left(z\right)$ (see Appendix) , $\chi_{\frac{1}{2}}$ satisfies as before the standard boundary conditions in $r=0$, i.e. $\chi_{\frac{1}{2}}\left(0,\varphi\right)=\chi_{\frac{1}{2}}'\left(0,\varphi\right)=0$, $A$ and $B$ are arbitrary complex numbers and $\eta\in[-\pi,\pi)$. Thus analogously as for  $\theta=\frac{1}{2}$ we also obtain  in the case of $\theta=0$ the one-parameter family of extensions.

We now discuss the bound states. As mentioned above, in the case with $\theta=0$ the time reversal symmetry does not imply any additional condition. Therefore, in this case our earlier observations concerning bound states hold true. On the other hand, in the case $\theta=\frac{1}{2}$ the additional condition $\eta=\rho$ reduces possible spectrum of negative energy states. Namely, we have in this case doubly degenerate energy level:
\begin{equation}\label{a44}
E\equiv E_0=E_{-1}=-\kappa\left(\frac{\cos\left(\frac{\eta}{2}+\frac{\pi}{8}\right)}{\cos\left(\frac{\eta}{2}-\frac{\pi}{8}\right)}\right)^2,\qquad \eta\in\left(-\frac{3}{4}\pi,\frac{3}{4}\pi\right]
\end{equation}

\begin{figure}[h]
		\begin{center}
		\includegraphics[scale=0.5]{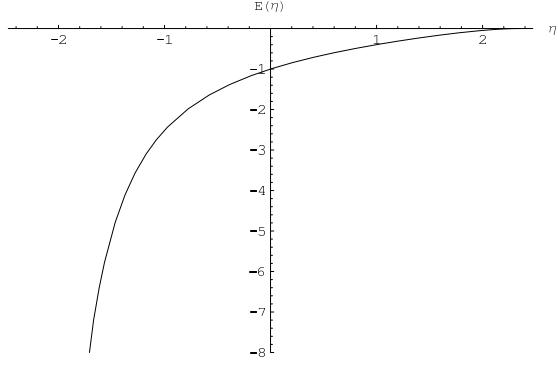}
		\end{center}
		\caption{\label{wykE}Energy of the bound state  \eqref{a44}. It should be noted that we have analogous situation as in Fig.\ref{wykE0} referring to the energy of the  bound state \eqref{a31}.}
\end{figure}

corresponding to the eigenfunctions
\begin{equation}\label{a45}
\begin{split}
\Psi_E^{\left(\frac{1}{2},0\right)}\left(r,\varphi\right)&=\sqrt{\frac{2\sqrt{2M|E|}}{\hbar}}\frac{e^{-\frac{r}{\hbar}\sqrt{2M|E|}}}{\sqrt{r}}e^{\frac{i\varphi}{2}},\\
\Psi_E^{\left(\frac{1}{2},-1\right)}\left(r,\varphi\right)&=\sqrt{\frac{2\sqrt{2M|E|}}{\hbar}}\frac{e^{-\frac{r}{\hbar}\sqrt{2M|E|}}}{\sqrt{r}}e^{-\frac{i\varphi}{2}}.
\end{split}
\end{equation}

\begin{figure}[h]
		\begin{center}
		\includegraphics[scale=0.7]{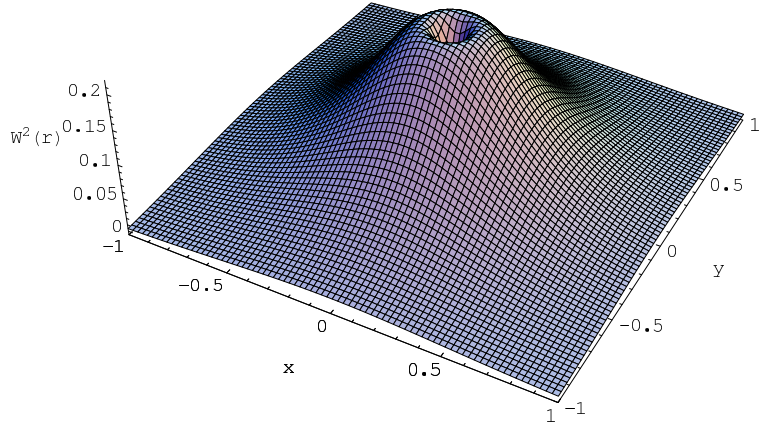}
		\end{center}
		\caption{\label{wykP}The probability density $W^2\left(r\right)=r|\Psi_E^{\left(\frac{1}{2},0\right)}\left(r\right)|^2$ (with respect to the measure $dr$), where $\Psi_E^{\left(\frac{1}{2},0\right)}\left(r\right)$ is given by  \eqref{a45}. The behavior of $W^2\left(r\right)$ at the origin is similar to $W_0^2\left(r\right)$ presented in Fig.\ref{wykK0}.}
\end{figure}

 If $\eta$ does not belong to the interval  $(-\frac{3}{4}\pi,\frac{3}{4}\pi]$ then  $\hat{H}$ does not have negative energies in its spectrum.

Finally, it should be noted that the form of the $SO\left(2\right)$ transformations \eqref{a3} is reduced by $\hat{T}$-symmetry to
\begin{equation}\label{a46}
\hat{U}\left(\alpha\right)\psi\left(r,\varphi\right)=\psi\left(r,\varphi+\alpha\right)
\end{equation}
i.e. we have  $\lambda=0$. Consequently, the momentum operator $\hat{J}=\hbar\hat{J}_0$ has the spectrum $\{\hbar m\}$, where $m\in{\bf Z}$ for $\theta=0$ and $\{\hbar \left(m+\frac{1}{2}\right)\}$, where $m\in{\bf Z}$, for $\theta=\frac{1}{2}$.

Finally, we comment on some statements of the paper \cite{Cirone}. First of all, the analysis of the two dimensional quantum mechanics by authors of  \cite{Cirone} is related to the pointed plane $\Dot{{\bf R}}^2$ rather than to the plane ${\bf R}^2$. Indeed, by using the polar coordinates they extract the origin from  the coordinate frame and effectively work with the pointed plane $\Dot{{\bf R}}^2$ which has completely different topology. Furthermore, in our opinion the formation of the bound states is, as was shown above, a consequence of the change of the  topology of the configuration space and it is not the result of the attractive centrifugal force. In fact, let us look at the radial Hamiltonian $\hat{H}_{\theta,m}$ \eqref{a17}. After unitary transformation $\Psi\left(r\right)\rightarrow\frac{u\left(r\right)}{\sqrt{r}}$ mapping $ L^2\left({\bf R}^2,rdr\right)$ into $ L^2\left({\bf R}^2,dr\right)$ the eigenvalue equation \eqref{a28} takes the form
\begin{equation}
\frac{\hbar^2}{2M}\left(\frac{\partial^2}{\partial r^2}-\frac{\left(\theta+m\right)^2-\frac{1}{4}}{r^2}\right)u_E^{\left(\theta,m\right)}\left(r\right)=|E|u_E^{\left(\theta,m\right)}\left(r\right)
\end{equation}

In  the case of the bound state discussed in \cite{Cirone} we have $\theta=0$ and $m=0$ (i.e s-wave) and  the centrifugal potential is indeed attractive. Nevertheless for $\frac{1}{2}\leq \theta<1$ and $m=0$ (or $0<\theta\leq\frac{1}{2}$ and  $m=-1$)  this effective potential is evidently repulsive or vanish in the $\hat{T}$-symmetric case with $\theta=\frac{1}{2}$, nevertheless the bound states can exists also in these cases as is evident from our discussion in section \ref{VI}. Moreover, in these cases the angular momentum does not vanish (it equals $\hbar\theta$ for $m=0$ or $\left(\theta-1\right)\hbar$ for $m=-1$, respectively) .

We remark that an advantage of the method of the self-adjoint extensions of symmetric operators applied in this work  in comparison with  the approach based on the formal Dirac distribution potential (Fermi pseudopotential) \cite{Albeverio,Wodkiewicz}  is that we can more naturally interprete the extension parameters as related to the boundary conditions specifying what happens in the extracted point. On the other hand, we would like to stress once more that the theory of self-adjoint extensions applied herein is the most adequate tool for the study of such subtle problems as quantum mechanics on a pointed plane.

\appendix
\section*{Appendix}

The general solution to the equation \cite{Erdei} 
\begin{equation*}\label{3.4}
\frac{\partial^2}{\partial r^2}\chi\left(r\right)+ \frac{1}{r}\frac{\partial}{\partial r}\chi\left(r\right)-\left(1+\frac{\mu^2}{r^2}\right)\chi\left(r\right)=0
\end{equation*}
can be written in the form

\begin{equation*}\label{3.5}
\chi\left(r\right)=C_1I_\mu\left(r\right)+C_2K_\mu\left(r\right),
\end{equation*}
where $I_\mu$ and $K_\mu$ are the modified Bessel and MacDonald functions, respectuvely.

\subsection*{Asymptotic behavior of $I_\mu$ and $K_\mu$ functions}
\begin{itemize}
\item {$\mu\neq 0$}

for $|z| \rightarrow 0$  we have
\[I_\mu\left(z\right)\sim \frac{1}{\Gamma\left(\mu+1\right)}\left(\frac{z}{2}\right)^\mu,\]
\[K_\mu\left(z\right)\sim \frac{1}{2}\Gamma\left(\mu\right)\left(\frac{z}{2}\right)^{-|\mu|},\]
for  $|z|\rightarrow\infty$ the asymptotic formulas can be written as
\[I_\mu\left(z\right)\sim \frac{e^z}{\sqrt{2\pi z}},\]
\[K_\mu\left(z\right)\sim \sqrt{\frac{\pi}{2 z}}e^{-z},\]

\item {$\mu=0$ }

 for $|z|\rightarrow 0$ we have 
\begin{equation*}
K_0\left(z\right)\sim\ln\frac{2}{z},
\end{equation*}

for  $|z|\rightarrow\infty$ the asymptotic relations can be written in the form
\[I_0\left(z\right)\sim \frac{e^z}{\sqrt{2\pi z}},\]

\[K_0\left(z\right)\sim \sqrt{\frac{\pi}{2 z}}e^{-z},\]

\end{itemize}

\subsection*{Some useful identities for $K_\mu$}
\[K_\mu\left(z\right)=K_{-\mu}\left(z\right),\]
\[\left({K_\mu\left(z\right)}\right)^*=K_{\mu}\left(z^*\right),\qquad \text{where } \mu\in{\bf R},\]
\[-\frac{2\mu}{z}K_\mu\left(z\right)=K_{\mu-1}\left(z\right)-K_{\mu+1}\left(z\right),\]
\[-2K_\mu'\left(z\right)=K_{\mu-1}\left(z\right)+K_{\mu+1}\left(z\right),\]
\[\frac{d}{dz}\left(z^\mu K_\mu\left(z\right)\right)=-z^\mu K_{\mu-1}\left(z\right),\]
\[\frac{d}{dz}\left(z^{-\mu}K_\mu\left(z\right)\right)=-z^{-\mu}K_{\mu+1}\left(z\right),\]
\[K_{\frac{1}{2}}\left(z\right)=\sqrt{\frac{\pi}{2z}}e^{-z},\]
\[\int_0^\infty xK_\mu\left(a x\right)K_\mu\left(b x\right)dx=\frac{\pi \left(ab\right)^{-\mu}\left(a^{2\mu}-b^{2\mu}\right)}{2\sin\left(\mu\pi\right)\left(a^2-b^2\right)},\]
where $\text{Re}\left(a+b\right)>0$ and $|\text{Re}\mu|<1$. In the limit $\mu=0$ the above formula takes the form
\[\int_0^\infty xK_0\left(a x\right)K_0\left(b x\right)dx=\frac{\ln{a}-\ln{b}}{a^2-b^2}.\]

\end{document}